\documentclass[11pt]{article}

\newcommand{\be}{\begin{equation}}
\newcommand{\ee}{\end{equation}}
\newcommand{\bea}{\begin{eqnarray}}
\newcommand{\eea}{\end{eqnarray}}
\newcommand{\sn}{{\rm sn}}

\newcommand{\dn}{{\rm dn}}
\newcommand{\cn}{{\rm cn}}
\newcommand{\sech}{{\rm sech}}

\begin{document}
\vspace{.5in}
\begin{center}
{\LARGE{\bf Domain Wall and Periodic Solutions of Coupled $\phi^6$ and  
Coupled $\phi^6$-$\phi^4$ Models}}
\end{center}

\vspace{.3in}
\begin{center}
{\LARGE{\bf Avinash Khare}} \\
{Institute for Physics, Bhubaneswar, Orissa 751005, India}
\end{center}

\begin{center}
{\LARGE{\bf Avadh Saxena}} \\
{Theoretical Division and Center for Nonlinear Studies, Los
Alamos National Lab, Los Alamos, NM 87545, USA}
\end{center}

\vspace{.9in}
{\bf {Abstract:}}

We obtain several higher order periodic solutions of a Coupled $\phi^6$ 
model in terms of Lam\'e polynomials of order one and two. These 
solutions are unusual in the sense that while they are the solutions of 
the coupled problem, they are not the solutions of the uncoupled problem. 
We also obtain exact solutions of coupled $\phi^6$-$\phi^4$ models, both 
when the $\phi^4$ potential corresponds to a first order (asymmetric 
double well) or a second order (symmetric double well) transition.

\newpage

\section{Introduction}

Coupled triple well or $\phi^6$ models \cite{sbh, lithium} arise in the 
context of many first order structural phase transitions.  There exist 
analogous coupled models in field theoretical contexts \cite{bazeia,santos, 
lou}.  Specifically, when a first order transition is driven by two primary 
order parameters, the free energy should be expanded to sixth order in both 
order parameters with a bi-quadratic (or possibly other symmetry allowed) 
coupling.  In a recent publication we obtained a large number of periodic 
solutions of a coupled $\phi^6$ model with bi-quadratic coupling \cite{ks1}. 
All these solutions had the feature that in the uncoupled limit, they reduce 
to the well known solutions of the corresponding uncoupled $\phi^6$ problem. 
The purpose of this paper is to point out that this coupled model has truly 
novel solutions in terms of Lam\'e polynomials of order one and two 
\cite{gr,finkel}, provided we add (symmetry allowed) quartic-quadratic and 
quadratic-quartic couplings to the coupled $\phi^6$ model considered earlier 
\cite{ks1}. It may be noted that these solutions exist only because of the 
coupling between the two fields (up to sixth order). In other words, while 
the Lam\'e polynomials of order one and two are the solutions of the coupled 
problem, they are not the solutions of the uncoupled problem. 

In this context, we note that in recent publications we have obtained a 
large number of periodic solutions, in terms of Lam\'e polynomials of order 
one and two, of coupled $\phi^4$ problems, both when the $\phi^4$ potential 
corresponds to the second as well as to the first order transition 
\cite{ks2,ks3,ks4}. The obvious question then is whether one can also obtain 
exact solutions of the coupled $\phi^6$-$\phi^4$ problems, both when the 
$\phi^4$ potential corresponds to the second as well as to the first order 
transition.  Interestingly, examples of these situations occur in condensed 
matter.  The face-centered cubic to monoclinic transition in Pu involves 
an intermediate phase hexagonal to monoclinic transition with a free energy 
modeled by the coupled $\phi^6$-asymmetric $\phi^4$ model \cite{pu}. 
Similarly, the coexistence of face-centered cubic, body-centered cubic and 
hexagonal close-packed structures in cobalt \cite{cobalt} as well as Fe and 
Tl \cite{dmit} is modeled by the same free energy.  An example of coupled 
$\phi^6$-symmetric $\phi^4$ model with a bi-quadratic coupling is the 
triggered ferroelectric transition \cite{hola,gufan}. Thus, another 
important purpose of this paper is to obtain exact solutions of the coupled 
$\phi^6$-$\phi^4$ models with bi-quadratic coupling, both when the $\phi^4$ 
potential corresponds to the first as well as to the second order transition.  

The plan of the paper is the following. In Sec. II we provide novel periodic 
as well as the corresponding hyperbolic solutions in terms of Lam\'e 
polynomials of {\it order one} for the coupled $\phi^6$ model with an explicit 
bi-quadratic as well as quadratic-quartic and quartic-quadratic couplings.    
In Sec. III we provide novel periodic as well as the corresponding hyperbolic 
solutions in terms of Lam\'e polynomials of {\it order two} for the same 
coupled $\phi^6$ model.  In Sec. IV we provide the exact solutions of the 
coupled $\phi^6-\phi^4$ problem with bi-quadratic coupling in case the 
$\phi^4$ potential corresponds to either the first or the second order 
transition.  Finally, in Sec. V we conclude with summary and possible 
extensions.

\section{The Coupled $\phi^6$ Model and Solutions in Terms of Lam\'e
Polynomials of Order One}

In \cite{ks1} we had considered the following coupled 
$\phi^6$ model, with a bi-quadratic coupling, in one dimension with 
the potential
\be\label{2.1}
V(\phi,\psi)=\left(\frac{a_1}{2} \phi^2 -\frac{b_1}{4}\phi^4 
+\frac{c_1}{6}\phi^6\right)  
+\left(\frac{a_2}{2} \psi^2 -\frac{b_2}{4}\psi^4 +\frac{c_2}{6}\psi^6 
\right) +\frac{d}{2}\phi^2 \psi^2\,. 
\ee
We now show that in case we add the following quartic-quadratic and 
quadratic-quartic coupling terms 
\be\label{2.2}
V'=\frac{e}{4}\phi^4\psi^2+\frac{f}{2}\phi^2\psi^4\,,
\ee
to the potential (\ref{2.1}), then in addition to the solutions
obtained in \cite{ks1}, there exist truly novel solutions in terms
of Lam\'e polynomials of order one and two to this coupled problem. 
Here $a_{1,2},b_{1,2},c_{1,2}$, $d$, $e$ and $f$ are material (or system) 
dependentparameters; $\phi$ and $\psi$ are scalar fields. From stability
considerations we shall always take $c_1,c_2>0$. Further, since we are
interested in a model for first order transition, we shall take
$b_1,b_2>0$. As far as $a_1,a_2$ are concerned, their sign is arbitrary 
and the shape of the potential depends on the ratio $b_1^2/4a_1c_1$ and
$b_2^2/4a_2c_2$. In particular, in the decoupled limit (i.e. $d=e=f=0$), it 
is easily shown that as long as $4a_1c_1>b_1^2$, the potential has a minimum 
at $\phi=0$ \cite{bk,ss}. 
In the case $4a_1c_1=b_1^2$, apart from the minimum at 
$\phi=0$ one now has points of inflection at $\phi^2=b_1/2c_1$. As $a_1$ 
decreases further so that $4a_1c_1 < b_1^2 < (16/3)a_1c_1$, one finds that 
while $\phi=0$ is still the absolute minimum, one now has two local minima 
and two maxima at \be\label{2.xx}
\phi_{min}^2=\frac{b_1+\sqrt{b_1^2-4a_1c_1}}{2c_1}\,,~~
\phi_{max}^2=\frac{b_1-\sqrt{b_1^2-4a_1c_1}}{2c_1}\,.
\ee
At the special value of $(16/3)a_1c_1=b_1^2$ one has three degenerate
minima at $\phi=0$ and at $\phi_{min}^2$ as given by Eq. (\ref{2.xx}); 
for relevant figures, see \cite{bk,ss}. This is the point of 
first order transition.  As $a_1$ decreases further so that 
$0<(16/3)a_1c_1<b_1^2$, then the roles of the minima are reversed, now 
$\phi=0$ is the local minimum while $\phi_{min}^2$ as given by Eq. 
(\ref{2.xx}) are the two degenerate absolute minima and $\phi_{max}^2$ are 
the two maxima. Finally as $a_1\le 0$, the potential has two absolute 
minima at $\phi_{min}^2$ as given by Eq. (\ref{2.xx}) while $\phi=0$ is 
now the sole maximum.  This picture continues to persist, for arbitrarily 
large and negative $a_1$. Throughout this paper, we shall term the point 
where $b_1^2=(16/3)a_1c_1$ as the point of first order transition, i.e. 
the point with $T=T_c$. On the other hand, the point where local structure 
(i.e. local minima) starts growing with decreasing temperature, i.e. 
$b_1^2=4a_1c_1$, will be termed as $T=T_p$. Note that for $T>T_p$ there is 
only a global minimum at $\phi=0$ and there are no other extrema. Thus the 
region where $b_1^2>(16/3)a_1c_1$ corresponds to $T<T_c$ while the region 
with $4a_1c_1 <b_1^2<(16/3)a_1c_1$ corresponds to $T_c<T<T_p$.
A similar analysis is also true for the potential in $\psi$
with $a_1,b_1,c_1$ being replaced by $a_2,b_2,c_2$, respectively.

The (static) equations of motion which follow from Eqs. (\ref{2.1}) 
and (\ref{2.2}) are
\bea\label{2.3}
&&\frac{d^2 \phi}{dx^2}= a_1 \phi -b_1\phi^3 +c_1 \phi^5 +d\phi \psi^2
+e\phi^3 \psi^2 +f \phi \psi^4\,,  \nonumber \\
&&\frac{d^2 \psi}{dx^2}= a_2 \psi -b_2\psi^3 +c_2 \psi^5 +d\psi \phi^2
+\frac{e}{2}\phi^4 \psi+f \phi^2 \psi^3\,. 
\eea
These coupled equations have thirteen distinct periodic (elliptic function) solutions, i.e. five ``bright-bright'', three ``bright-dark''
and five ``dark-dark'' solutions which have already been discussed in
\cite{ks1} (in case $e=f=0$). In particular, there are five
solutions below the transition temperature $T_c$, four at $T_c$, one
above $T_c$ (i.e. $T_c < T < T_p$), and three in the mixed phase in the sense that while one of the field is above $T_c$, the other one is 
below $T_c$. The latter situation is akin to the one found in multiferroic 
materials where one transition (i.e. antiferromagnetic) takes place at
a higher temperature than the other transition (e.g. ferroelectric) 
or vice versa \cite{multif}.  It is worth pointing 
out that in turn in the single soliton limit, these lead to {\it eight} 
distinct coupled (hyperbolic) soliton solutions.  In particular, 
one obtains three 
solutions below the transition temperature $T_c$, two at $T_c$, one above 
$T_c$ and two in the mixed phase in the sense that while one of the field 
is above $T_c$, the other one is below $T_c$.

We now show that apart from these solutions, we also have rather unusual
solutions in terms of Lam\'e polynomials of order one (and two discussed in
the next section) which we now
discuss one by one. Since there are three Lam\'e polynomials of order one 
(i.e. $\sn,\cn,\dn$) and since the field equations are essentially symmetric 
in $\phi$ and $\psi$, we expect six independent solutions to the
coupled field equations in terms of Lam\'e polynomials of order one.  
In particular, we first show that there are three periodic
bright-bright, two periodic dark-bright and one periodic dark-dark soliton
solutions in terms of Lam\'e polynomials of order one,
which in turn lead to one bright-bright, one dark-dark and one dark-bright
hyperbolic soliton solution. 

\subsection{Solution I}

We look for the most general solutions to the coupled Eqs. (\ref{2.3}) 
in terms of the Jacobi elliptic functions $\sn(x,m)$, $\cn(x,m)$ and 
$\dn(x,m)$ 
\cite{gr} where the modulus $m \equiv k^2$.  It is easily shown that
\be\label{2.4}
\phi=A\sn(Dx+x_0,m)\,,~~\psi=B\sn(Dx+x_0,m)\,,
\ee
is an exact dark-dark periodic solution to the coupled Eqs. (\ref{2.3}) 
provided the following six coupled equations are satisfied
\be\label{2.5}
a_1=-(1+m)D^2\,,
\ee
\be\label{2.6}
-b_1A^2+dB^2=2mD^2\,,
\ee
\be\label{2.7}
c_1^4+eA^2B^2+\frac{f}{2}B^4=0\,,
\ee
\be\label{2.8}
a_2=-(1+m)D^2\,,
\ee
\be\label{2.9}
-b_2B^2+dA^2=2mD^2\,,
\ee
\be\label{2.10}
c_2B^4+fA^2B^2+\frac{e}{2}A^4=0\,.
\ee
Here, $A$ and $B$ denote the amplitudes of the kink lattice, $D$ is an
inverse characteristic length while $x_0$ is the (arbitrary) location of
the kink. Three of these equations determine the three unknowns $A,B,D$
while the other three equations give three constraints between the nine
parameters $a_{1,2},b_{1,2},c_{1,2},d,e,f$. In particular, we find that the
solution exists only if $a_1<0,a_2<0,e<0,f<0$. We obtain
\be\label{2.11}
D^2=\frac{|a_1|}{(1+m)}\,,~~
A^2=\frac{2mD^2(d+b_2)}{d^2-b_1b_2}\,,~~B^2=\frac{(d+b_1)A^2}{(d+b_2)}\,,
\ee
while the three constraints are
\bea\label{2.12}
&&a_1=a_2<0\,,~~(4c_1c_2-|e||f|)(d+b_1)=2(e^2+2|f|c_1)(d+b_2)\,,
\nonumber \\
&&(4c_1c_2-|e||f|)^2=4(e^2+2|f|c_1)(f^2+2|e|c_2)\,.
\eea

In the limit of $m=1$, the periodic solution (\ref{2.4}) goes over to
the hyperbolic dark-dark soliton solution
\be\label{2.13}
\phi=A\tanh(Dx+x_0)\,,~~\psi=B\tanh(Dx+x_0)\,,
\ee
provided the constraints (\ref{2.11}) and (\ref{2.12}) with $m=1$ are
satisfied.

\subsection{Solution II}

It is easy to show that 
\be\label{2.14}
\phi=A\cn(Dx+x_0,m)\,,~~\psi=B\cn(Dx+x_0,m)\,,
\ee
is an exact bright-bright periodic solution to the coupled Eqs. (\ref{2.3}) 
provided six coupled equations similar to Eqs. (\ref{2.5}) to (\ref{2.10})
are satisfied. 
Three of these equations again determine the three unknowns $A,B,D$
while the other three equations give three constraints between the nine
parameters $a_{1,2},b_{1,2},c_{1,2},d,e,f$. In particular, we obtain
\be\label{2.15}
D^2=\frac{a_1}{(2m-1)}\,,~~a_1=a_2\,,~~
A^2=\frac{2m(d+b_2)D^2}{(b_1b_2-d^2)}\,,~~B^2=\frac{(b_1+d) A^2}{(b_2+d)}\,,
\ee
while the remaining two constraints are again given by Eq. (\ref{2.12}).
Note that $a_1=a_2> (<)0$ if $m> (<) 1/2$. 

In the limit of $m=1$, the periodic solution (\ref{2.14}) goes over to
the hyperbolic bright-bright solution
\be\label{2.16}
\phi=A\sech(Dx+x_0)\,,~~\psi=B\sech(Dx+x_0)\,,
\ee
provided the constraints (\ref{2.12}) and (\ref{2.15}) with $m=1$ are
satisfied.

\subsection{Solution III}

Yet another bright-bright periodic soliton solution is
\be\label{2.17}
\phi=A\dn(Dx+x_0,m)\,,~~\psi=B\dn(Dx+x_0,m)\,,
\ee
provided six coupled equations similar to Eqs. (\ref{2.5}) to (\ref{2.10})
are satisfied. 
Three of these equations determine the three unknowns $A,B,D$
while the other three equations give three constraints between the nine
parameters $a_{1,2},b_{1,2},c_{1,2},d,e,f$. In particular, we obtain
\be\label{2.18}
D^2=\frac{a_1}{(2-m)}\,,~~a_1=a_2>0\,,~~
A^2=\frac{2(d+b_2)D^2}{(b_1b_2-d^2)}\,,~~B^2=\frac{(b_1+d) A^2}{(b_2+d)}\,,
\ee
while the remaining two constraints are again given by Eq. (\ref{2.12}).

In the limit of $m=1$, the periodic solution (\ref{2.17}) again goes over to
the hyperbolic bright-bright soliton solution (\ref{2.16}).

Note that while for the solution (\ref{2.4}), $d^2>b_1b_2$,
for the solutions (\ref{2.14}) and (\ref{2.17}), its the other way around, 
i.e. $d^2<b_1b_2$. 

\subsection{Solution IV}

Yet another bright-bright periodic soliton solution is
\be\label{2.19}
\phi=A\sqrt{m}\cn(Dx+x_0,m)\,,~~\psi=B\dn(Dx+x_0,m)\,,
\ee
provided the following six coupled equations are satisfied
\be\label{2.20}
a_1+\frac{f}{2}(1-m)^2B^4+d(1-m)B^2=(2m-1)D^2\,,
\ee
\be\label{2.21}
b_1A^2+dB^2-(1-m)eA^2B^2-(1-m)fB^2=2D^2\,,
\ee
\be\label{2.22}
c_1A^4+eA^2B^2+\frac{f}{2}B^4=0\,,
\ee
\be\label{2.23}
a_2+\frac{e}{2}(1-m)^2 A^4-(1-m)dA^2=(2-m)D^2\,,
\ee
\be\label{2.24}
b_2B^2+dA^2+(1-m)fA^2B^2+(1-m)eA^4=2D^2\,,
\ee
\be\label{2.25}
c_2B^4+fA^2B^2+\frac{e}{2}A^4=0\,.
\ee
Three of these equations determine the three unknowns $A,B,D$
while the other three equations give three constraints between the nine
parameters $a_{1,2},b_{1,2},c_{1,2},d,e,f$. In particular, this solution is
also valid only if $e<0,f<0$. Further, two of the
relations are  given by 
\be
(4c_1c_2-|e||f|)^2=4(e^2+2|f|c_1)(f^2+2|e|c_2)\,,~~(4c_1c_2-|e||f|)A^2=
2(f^2+2|e|c_2)B^2\,.
\ee

In the limit of $m=1$, the periodic solution (\ref{2.18}) again goes over to
the hyperbolic bright-bright soliton solution (\ref{2.16}).

\subsection{Solution V}

In addition there are two dark-bright periodic soliton solutions. One of 
them is
\be\label{2.26}
\phi=A\sn(Dx+x_0,m)\,,~~\psi=B\cn(Dx+x_0,m)\,,
\ee
provided the following six coupled equations are satisfied
\be\label{2.27}
a_1+\frac{f}{2}B^4+dB^2=-(1+m)D^2\,,
\ee
\be\label{2.28}
-b_1A^2-dB^2+eA^2B^2-fB^4=2mD^2\,,
\ee
\be\label{2.29}
c_1A^4+\frac{f}{2}B^4=eA^2B^2\,,
\ee
\be\label{2.30}
a_2+\frac{e}{2} A^4+dA^2=(2m-1)D^2\,,
\ee
\be\label{2.31}
b_2B^2+dA^2-fA^2B^2+eA^4=2mD^2\,,
\ee
\be\label{2.32}
c_2B^4+\frac{e}{2}A^4=fA^2B^2\,.
\ee
Three of these equations determine the three unknowns $A,B,D$
while the other three equations give three constraints between the nine
parameters $a_{1,2},b_{1,2},c_{1,2},d,e,f$. For example, it is easily shown 
that
\be\label{2.33}
A^2=\frac{b_1\pm\sqrt{b_1^2-4a_1c_1-4(1-m)D^2c_1}}{2c_1}\,,~~
B^2=\frac{b_2\pm\sqrt{b_2^2-4a_2c_2-4D^2c_2}}{2c_2}\,.
\ee
Further, unlike the 
previous four solutions, this solution exists only if $e>0,f>0$ and two of the
constraints are given by 
\be\label{2.34}
(4c_1c_2-ef)A^2=2(2ec_2-f^2)B^2\,,~~(4c_1c_2-ef)^2=4(2ec_2-f^2)(2fc_1-e^2)\,, 
\ee
while the other two constraints are
\be\label{2.35}
B^2=\frac{b_1A^2-2a_1-2D^2}{d+eA^2}=
\frac{2a_2+2(1-m)D^2+dA^2}{b_2-fA^2}\,.
\ee

In the limit of $m=1$, the periodic solution (\ref{2.26}) goes over to
the hyperbolic dark-bright soliton solution 
\be\label{2.36}
\phi=A\tanh(Dx+x_0)\,,~~\psi=B\sech(Dx+x_0)\,,
\ee
satisfying the constraints (\ref{2.33}) to (\ref{2.35}) with $m=1$. 

\subsection{Solution VI}

Another dark-bright periodic soliton solution is given by 
\be\label{2.37}
\phi=A\sqrt{m}\sn(Dx+x_0,m)\,,~~\psi=B\dn(Dx+x_0,m)\,,
\ee
provided six coupled equations similar to (\ref{2.27}) to (\ref{2.32}) are 
satisfied.
Three of these equations determine the three unknowns $A,B,D$
while the other three equations give three constraints between the nine
parameters $a_{1,2},b_{1,2},c_{1,2},d,e,f$. For example, it is easily shown 
that two of the constraints are again given by Eq. (\ref{2.34}) while $A^2$
and $B^2$ are now given by
\be\label{2.38}
A^2=\frac{b_1\pm\sqrt{b_1^2-4a_1c_1+4(1-m)D^2c_1}}{2c_1}\,,~~
B^2=\frac{b_2\pm\sqrt{b_2^2-4a_2c_2-4mD^2c_2}}{2c_2}\,.
\ee
while the other two constraints are
\be\label{2.39}
B^2=\frac{b_1A^2-2a_1-2mD^2}{d+eA^2}=
\frac{2a_2-2(1-m)D^2+dA^2}{b_2-fA^2}\,.
\ee

In the limit of $m=1$, the periodic solution (\ref{2.37}) goes over to
the hyperbolic dark-bright soliton solution (\ref{2.36}).
 
\section{Solutions of Coupled $\phi^6$ Model In terms of Lam\'e Polynomials
of Order Two}

We now show that quite remarkably, the coupled model characterized by the
field Eqs. (\ref{2.3}) not only admits periodic solutions in terms
of Lam\'e polynomials of order one, but it also admits novel periodic
solutions in terms of Lam\'e polynomials of order two. It is worth reminding 
once again that neither Lam\'e polynomials of order one nor of order two
are solutions of the uncoupled $\phi^6$ problem. Since there are five Lam\'e
polynomials of order two, and since two of these are of the form 
$A\sn^2[D(x+x_0),m]+F$, and further, the two field Eqs. (\ref{2.3}) 
are symmetrical
in $\phi$ and $\psi$, in principle there could be ten solutions of  
order two. However, it turns out that only two of these are admitted by the
field Eqs. (\ref{2.3}) which we now discuss.
 
\subsection{Solution I}

It is easily shown that
\be\label{3.1}
\phi=A\sn^2(Dx+x_0,m)+F\,,~~\psi=B\sn(Dx+x_0,m)\cn(Dx+x_0,m)\,,
\ee
is an exact periodic solution to the coupled Eqs. (\ref{2.3}) 
provided the following eleven coupled equations are satisfied
\be\label{3.2}
a_1F-b_1F^3+c_1F^5=2AD^2\,,
\ee
\be\label{3.3}
a_1A-3b_1AF^2+5c_1AF^4+eB^2F^3+dB^2F=-4(1+m)AD^2\,,
\ee
\be\label{3.4}
-3b_1A^2F+10c_1A^2F^3+eB^2F^2(3A-F)+\frac{f}{2}B^4F+dB^2(A-F)=6mAD^2\,,
\ee
\be\label{3.5}
-b_1A^3+10c_1A^3F^2+3eAB^2F(A-F)+\frac{f}{2}B^4(A-2F)-dAB^2=0\,,
\ee
\be\label{3.6}
5c_1A^4F+eA^2B^2(A-3F)+\frac{f}{2}B^4(F-2A)=0\,,
\ee
\be\label{(3.7}
c_1A^4-eA^2B^2+\frac{f}{2}B^4=0\,.
\ee
\be\label{3.8}
a_2+\frac{e}{2}F^4+dF^2=-(4+m)D^2\,,
\ee
\be\label{3.9}
-b_2B^2+2eAF^3+fF^2B^2+2dAF=6mD^2\,,
\ee
\be\label{3.10}
b_2B^2+c_2B^4+3eA^2F^2+fB^2F(2A-F)+dA^2=0\,,
\ee
\be\label{3.11}
-2c_2B^4+2eA^3F+fAB^2(A-2F)=0\,,
\ee
\be\label{3.12}
c_2B^4-fA^2B^2+\frac{e}{2}A^4=0\,.
\ee
Four of these equations determine the four unknowns $A,B,D,F$
while the other equations give constraints between the nine
parameters $a_{1,2},b_{1,2},c_{1,2},d,e,f$. In particular, we find that the
solution exists only if 
\be\label{3.13}
eA^2=fB^2\,,~~e^3=8c_1^2c_2\,,~~f^3=8c_2^2c_1\,,
\ee
and further if $F \ne 0$. 

In the limit of $m=1$, the periodic solution (\ref{3.1}) goes over to
the hyperbolic solution
\be\label{3.14}
\phi=A\tanh^2(Dx+x_0)+F\,,~~\psi=B\tanh(Dx+x_0)\sech(Dx+x_0)\,,
\ee
provided the constraints (\ref{3.2}) to (\ref{3.12}) with $m=1$ are
satisfied. There is one special case when this solution takes a simpler
form, i.e. when $A=-F$, the solution is given by
\be\label{3.15}
\phi=-A\sech^2(Dx+x_0)\,,~~\psi=B\tanh(Dx+x_0)\sech(Dx+x_0)\,,
\ee
provided Eq. (\ref{3.13}) is satisfied and further
\bea\label{3.16}
&&D^2=\frac{a_1}{4}\,,~~a_1=4a_2>0\,,~~d=-b_2\,,~~b_2^2(f-e)=6a_2c_2e\,, 
\nonumber \\
&&B^2= \frac{6a_2}{b_2}\,,~~A^2=\frac{-b_1\pm\sqrt{b_1^2-6a_1c_1}}{2c_1}\,.
\eea
Thus in the $\phi$ variable, one is at $T<T_c^{I}$ since $b_1^2>6a_1c_1$.

\subsection{Solution II}

The other allowed solution is
\be\label{3.17}
\phi=A\sn^2(Dx+x_0,m)+F\,,~~\psi=B\sn(Dx+x_0,m)\dn(Dx+x_0,m)\,,
\ee
which is an exact periodic solution to the coupled Eqs. (\ref{2.3}) 
provided Eqs. (\ref{3.2}), (\ref{3.3}) and the following nine coupled equations are satisfied
\be\label{3.18}
-3b_1A^2F+10c_1A^2F^3+eB^2F^2(3A-mF)+\frac{f}{2}B^4F+dB^2(A-mF)=6mAD^2\,,
\ee
\be\label{3.19}
-b_1A^3+10c_1A^3F^2+3eAB^2F(A-mF)+\frac{f}{2}B^4(A-2mF)-dmAB^2=0\,,
\ee
\be\label{3.20}
5c_1A^4F+eA^2B^2(A-3mF)+\frac{mf}{2}B^4(mF-2A)=0\,,
\ee
\be\label{(3.21}
c_1A^4-emA^2B^2+\frac{f}{2}m^2B^4=0\,.
\ee
\be\label{3.22}
a_2+\frac{e}{2}F^4+dF^2=-(1+4m)D^2\,,
\ee
\be\label{3.23}
-b_2B^2+2eAF^3+fF^2B^2+2dAF=6mD^2\,,
\ee
\be\label{3.24}
b_2mB^2+c_2B^4+3eA^2F^2+fB^2F(2A-mF)+dA^2=0\,,
\ee
\be\label{3.25}
-2mc_2B^4+2eA^3F+fAB^2(A-2mF)=0\,,
\ee
\be\label{(3.26}
c_2m^2B^4-fmA^2B^2+\frac{e}{2}A^4=0\,.
\ee
Four of these equations determine the four unknowns $A,B,D,F$
while the other equations give constraints between the nine
parameters $a_{1,2},b_{1,2},c_{1,2},d,e,f$. In particular, we find that the
solution exists only if 
\be\label{3.27}
emA^2=fB^2\,,~~e^3=8c_1^2c_2\,,~~f^3=8c_2^2c_1\,,
\ee
and further if $F \ne 0$. 

In the limit of $m=1$, the periodic solution (\ref{3.17}) also goes over to
the hyperbolic solution (\ref{3.14}). 

\section{Coupled $\phi^6$-$\phi^4$ Model}

We now consider a coupled $\phi^6$-$\phi^4$ model with bi-quadratic coupling. 
In particular, we consider the model characterized by the potential
\be\label{4.1}
V(\phi,\psi)=\left(\frac{a_1}{2} \phi^2 -\frac{b_1}{4}\phi^4 
+\frac{c_1}{6}\phi^6\right)  
+\left(\frac{a_2}{2} \psi^2 +\frac{f_2}{3}\psi^3 +\frac{b_2}{4}\psi^4 
\right) +\frac{d}{2}\phi^2 \psi^2\,. 
\ee
This leads to the coupled field equations
\bea\label{4.2}
&&\frac{d^2 \phi}{dx^2}= a_1 \phi -b_1\phi^3 +c_1 \phi^5 +d\phi \psi^2\,, 
\nonumber \\
&&\frac{d^2 \psi}{dx^2}= a_2 \psi +f_2 \psi^2+b_2\psi^3 +d\psi \phi^2\,. 
\eea
From stability considerations we shall always take $c_1>0,b_2>0$. Note that in
case $f_2=0$, the model corresponds to the symmetric $\phi^4$ model with  
a second order transition, while as long as $f_2 \ne 0$, the model 
corresponds to a first order transition. We shall discuss the various
solutions both when $f_2 \ne 0$ as well as when $f_2=0$.

\subsection{Solution I}

It is easily shown that
\be\label{4.3}
\phi=A\sqrt{1\pm \sn(Dx+x_0,m)}\,,~~\psi=B\sn(Dx+x_0,m)+F\,,
\ee
is an exact periodic solution to the coupled Eqs. (\ref{4.2}) 
provided the following seven coupled equations are satisfied
\be\label{4.4}
a_1-b_1A^2+c_1A^4+dF^2=-\frac{D^2}{4}\,,
\ee
\be\label{4.5}
-b_1A^2+2c_1A^4\pm 2dBF=-\frac{mD^2}{2}\,,
\ee
\be\label{4.6}
c_1^4+dB^2=\frac{3mD^2}{4}\,,
\ee
\be\label{4.7}
[a_2+f_2F+b_2F^2+dA^2]F=0\,,
\ee
\be\label{4.8}
a_2B+2f_2FB+3b_2BF^2+dA^2(B\pm F)=-(1+m)BD^2\,,
\ee
\be\label{4.9}
[f_2B+3b_2BF\pm dA^2]B=0\,,
\ee
\be\label{4.10}
b_2B^2=2mD^2\,.
\ee
Here, $A$ and $B$ denote the amplitudes of the kink lattice, $D$ is an
inverse characteristic length, F is a constant while $x_0$ is the (arbitrary) 
location of
the kink. Four of these equations determine the four unknowns $A,B,D,F$
while the other three equations give three constraints between the seven
parameters $a_{1,2},b_{1,2},c_{1},f_2,d$. 

In the limit of $m=1$, the periodic solution (\ref{4.3}) goes over to
the hyperbolic solution
\be\label{4.11}
\phi=A\sqrt{1\pm \tanh(Dx+x_0)}\,,~~\psi=B\tanh(Dx+x_0)+F\,,
\ee
provided Eqs. (\ref{4.4}) to (\ref{4.10}) with $m=1$ are satisfied.

Several comments are in order at this stage.

\begin{enumerate}

\item The solution (\ref{4.3}) continues to exist even if $f_2=0$, i.e. in the 
symmetric $\phi^4$ case.

\item Solution (\ref{4.3}) continues to exist if $F=0$. However, no solution
exists in case $F=f_2=0$ as in that case $d$ is also forced to be zero. 

\item In case $B=\pm F$ then the solution exists only at $m=1$. In 
particular it is easily shown that 
\be\label{4.12}
\phi=A\sqrt{1\pm \tanh(Dx+x_0)}\,,~~\psi=F[1\pm\tanh(Dx+x_0)]\,,
\ee
is an exact solution to field Eqs. (\ref{4.2}) provided
\be\label{4.13}
D^2=a_1\,,~~a_2=4a_1>0\,,~~A^2=\frac{2a_1}{b_1}\,,~~B^2=\frac{a_2}{2b_2}\,,
\ee
and further
\be\label{4.14}
d=\frac{b_2a_1(3b_1^2-16a_1c_1)}{2a_2b_1^2}\,,~~
f_2 B=-6a_1-\frac{b_2a^2_1(3b_1^2-16a_1c_1)}{a_2b_1^3}\,.
\ee
In the special case of $f_2=0$, this solution continues to exist provided 
$f=-3b_1$ and hence $(16a_1c_1-3b_1^2)a_1b_2=6a_2b_1^3$.

\item On the other hand, the solution
\be\label{4.15}
\phi=A\sqrt{1\pm \tanh(Dx+x_0)}\,,~~\psi=F[1\mp \tanh(Dx+x_0)]\,,
\ee
is an exact solution to field Eqs. (\ref{4.2}) provided
\bea\label{4.16}
&&D^2+a_1=b_1A^2\,,~~2D^2+a_1=4c_1A^4\,,~~D^2-a_1=4dB^2\,, \nonumber \\
&&8D^2+a_2=2f_2B\,,~~4D^2-a_2=2dA^2\,,~~2D^2=b_2B^2\,.
\eea
These relations imply that $4c_1A^2=b_1\pm\sqrt{b_1^2-4a_1c_1}$. 
In the special case of $f_2=0$, this solution exists provided
$a_2=-8D^2\,,dA^2=6D^2$ while the other four relations are as given by
Eq. (\ref{4.16}).

\end{enumerate}

\subsection{Solution II}

We now present three solutions which are only valid at $m=1$, i.e. in the 
hyperbolic limit. For example, it is easy to show that 
\be\label{4.17}
\phi=\frac{A\sn(Dx+x_0,m)}{\sqrt{1-F\sn^2(Dx+x_0,m)}}\,,~~
\psi=\frac{B\cn^2(Dx+x_0,m)}{[1-F\sn^2(Dx+x_0,m)]}\,,
\ee
is an exact solution to the coupled Eqs. (\ref{4.2}) only if $m=1$, 
and if
\bea\label{4.18}
&&a_1+dB^2=(3F-2)D^2\,,~~b_1A^2=2(1-F)(D^2+a_1)\,,~~c_1A^4=(1-F)^2(2D^2+a_1)\,,
\nonumber \\
&&a_2+f_2B=-2(1+3F)D^2\,,~~dA^2=(1-F)[6(1+F)D^2-f_2B]\,,~~b_2B^2=8D^2F\,.
\eea
From here it follows that $(b_1^2-4a_1c_1)A^4=4(1-F)^2D^4>0$.
Note that at $m=1$, the solution (\ref{4.17}) can be rewritten as
\be\label{4.19}
\phi=\frac{A\tanh(Dx+x_0)}{\sqrt{1-F\tanh^2(Dx+x_0)}}\,,~~
\psi=\frac{B\sech^2(Dx+x_0)}{[1-F\tanh^2(Dx+x_0)]}\,,
\ee
Note also that this solution continues to hold good even if $f_2=0$.

\subsection{Solution III}

Another solution, which is only valid at $m=1$ is given by
\be\label{4.20}
\phi=\frac{A\sech(Dx+x_0)}{\sqrt{1-F\tanh^2(Dx+x_0)}}\,,~~
\psi=\frac{B\sech^2(Dx+x_0)}{[1-F\tanh^2(Dx+x_0)]}\,.
\ee
This is an exact solution to the coupled Eqs. (\ref{4.2}) provided 
\bea\label{4.21}
&&a_1=D^2\,,~~a_2=4a_1\,,~~b_1A^2=2(1+F)D^2\,,~~b_2B^2=8FD^2\,, \nonumber \\
&&c_1A^4+dB^2=3FD^2\,,~~dA^2+f_2B=-6(1+F)D^2\,.
\eea 
Note that this solution continues to hold good even if $f_2=0$.
 
\subsection{Solution IV}

Yet another solution, which is only valid at $m=1$ is given by
\be\label{4.22}
\phi=\frac{A}{\sqrt{1-F\tanh^2(Dx+x_0)}}\,,~~
\psi=\frac{B\sech^2(Dx+x_0)}{[1-F\tanh^2(Dx+x_0)]}\,.
\ee
This is an exact solution to the coupled Eqs. (\ref{4.2}) provided 
\bea\label{4.23}
&&b_1A^2=2(1-F)(D^2+a_1)\,,~~dB^2+a_1F^2=(3-2F)FD^2\,,~~
c_1A^4=(2D^2+a_1)(1-F)^2\,, \nonumber \\
&&a_2F+f_2B=-2(3+F)D^2\,,~~b_2B^2=8FD^2\,,~~dA^2=(1-F)(4D^2-a_2)\,.
\eea 
From here it follows that $(b_1^2-4a_1c_1)A^4=4(1-F)^2D^4>0$.
Note that this solution continues to hold good even if $f_2=0$.

It is worth noting that for the three solutions as given by Eqs. (\ref{4.19}),
(\ref{4.20}) and (\ref{4.22}), while $\phi$ continues to be a solution of
the uncoupled $\phi^6$ field theory, in neither of these three cases, 
$\psi$ is an exact solution of either the symmetric or the asymmetric, 
uncoupled $\psi^4$ problem.

\section{Conclusions}

In this paper we have shown that the Lam\'e polynomials of order one and
two are periodic solutions of a coupled $\phi^6$ problem. These are novel
solutions in the sense that while they are the solutions of the coupled
$\phi^6$ problem, they are not the solutions of the corresponding uncoupled 
problems.
In particular, we have obtained six solutions in terms of Lam\'e polynomials 
of order one and two solutions in terms of Lam\'e polynomials of order two.
These results are applicable to both the structural phase transitions 
\cite{sbh,lithium} and field theoretic contexts \cite{bazeia,santos,lou}.   

We have also obtained four solutions of the coupled $\phi^6-\phi^4$ problem,
both when the $\phi^4$ potential corresponds to a first order as well as
a second order transition. Note that while the solutions of the coupled 
problem are also the solutions of the uncoupled 
$\phi^6$ problem, but they are not the solutions of either the symmetric 
or the asymmetric uncoupled $\phi^4$ problems. These solutions 
are also useful in understanding coexistence of different crystalline 
structures in elements \cite{pu,cobalt,dmit} and ferroelectrics 
\cite{hola,gufan}.  

It will be interesting to obtain solutions of few other coupled field 
theories and with couplings that are not bi-quadratic.  An example of 
a coupled model with linear-quadratic coupling occurs in the context of 
isostructural transitions \cite{kutin}.  It is conceivable that in some 
cases a linear-cubic coupling may be symmetry allowed.  

\section{Acknowledgment}
This work was supported in part by the U.S. Department of Energy.

\end{document}